\documentclass[]{elsart5p}
\usepackage{graphicx}
\usepackage{enumerate}
\usepackage{amssymb}
\usepackage[usenames]{color}		% for \textcolor{red} ?
\usepackage{multirow}	% in tables
\usepackage{rotating}	% rotated labels, e.g., in tables
\usepackage{booktabs}	% book style tables
\usepackage{url}

\journal{Nuclear Instruments and Methods}

\def\\clight{\(c\,\)}

\begin{document}
\begin{frontmatter}



\title{Charged Kaon Mass Measurement using the Cherenkov Effect}


%Collaboration list
\author[IU]{N.~Graf}
\author[HU]{A.~Lebedev}
\author[UM,MI]{R.~J.~Abrams}
\author[UI]{U.~Akgun}
\author[UI]{G.~Aydin}
\author[FNAL]{W.~Baker}
\author[LLNL]{P.~D.~Barnes, Jr.~}
\author[USC]{T.~Bergfeld}
\author[FNAL]{L.~Beverly}
\author[PU]{A.~Bujak}
\author[FNAL]{D.~Carey}
\author[UV]{C.~Dukes}
\author[UI]{F.~Duru}
\author[HU]{G.J.~Feldman}
\author[USC]{A.~Godley}
\author[UI,AA]{E.~G\"{u}lmez}
\author[UI]{Y.O.~G\"{u}nayd{\i}n}
\author[UM]{H.~R.~Gustafson}
\author[PU]{L.~Gutay}
\author[LLNL]{E.~Hartouni}
\author[IIT]{P.~Hanlet}
\author[FNAL]{S.~Hansen}
\author[LLNL]{M.~Heffner}
\author[FNAL]{C.~Johnstone}
\author[IIT]{D.~Kaplan}
\author[IIT]{O.~Kamaev}
\author[FNAL]{J.~Kilmer}
\author[LLNL]{J.~Klay}
\author[FNAL]{M.~Kostin}
\author[LLNL]{D.~Lange}
\author[USC]{J.~Ling}
\author[UM]{M.~J.~Longo}
\author[UV]{L.~C.~Lu}
\author[UV]{C.~Materniak}
\author[IU]{M.D.~Messier}
\author[FNAL,CW]{H.~Meyer}
\author[PU]{D.~E.~Miller}
\author[USC]{S.~R.~Mishra}
\author[UV]{K.~Nelson}
\author[UM]{T.~Nigmanov}
\author[UV]{A.~Norman}
\author[UI]{Y.~Onel}
\author[IU]{J.~M.~Paley}
\author[UM]{H.~K.~Park}
\author[UI]{A.~Penzo}
\author[UCB]{R.~J.~Peterson}
\author[FNAL]{R.~Raja}
\author[UM]{D.~Rajaram}
\author[IIT]{D.~Ratnikov}
\author[USC]{C.~Rosenfeld}
\author[IIT]{H.~Rubin}
\author[HU]{S.~Seun}
\author[IIT,CW]{N.~Solomey}
\author[LLNL]{R.~Soltz}
\author[ELM]{E.~Swallow}
\author[FNAL]{R.~Schmitt}
\author[UM]{P.~Subbarao}
\author[IIT]{Y.~Torun}
\author[FNAL]{T.~E.~Tope}
\author[USC]{K.~Wilson}
\author[LLNL]{D.~Wright}
\author[USC]{K.~Wu}

\address[BNL]{Brookhaven National Laboratory, Upton NY 11973}
\address[ELM]{Elmhurst College, Elmhurst, IL 60126}
\address[FNAL]{Fermi National Accelerator Laboratory, Batavia, IL 60510}
\address[HU]{Harvard University, Cambridge, MA 02138}
\address[IIT]{Illinois Institute of Technology, Chicago, IL 60616}
\address[IU]{Indiana University, Bloomington, IN 47403}
\address[LLNL]{Lawrence Livermore National Laboratory, Livermore, CA 94550}
\address[PU]{Purdue Universty, West Lafayette, IN 47907}
\address[UCB]{University of Colorado, Boulder, CO 80309}
\address[UI]{University of Iowa, Iowa City, IA 52242}
\address[UM]{University of Michigan, Ann Arbor, MI 48109}
\address[USC]{University of South Carolina, Columbia, SC 29201}
\address[UV]{University of Virginia, Charlottesville, VA 22904}
\address[MI]{Currently at Muons, Inc., Batavia, IL 60510}
\address[AA]{Also at Bogazici University, Istanbul, Turkey}
\address[CW]{Currently at Wichita State University, Wichita, KS 67260}
\begin{abstract}
The two most recent and precise measurements of the charged kaon mass use X-rays from kaonic atoms and report uncertainties of 14~ppm and 22~ppm yet differ from each other by 122~ppm. We describe the possibility of an independent mass measurement using the measurement of Cherenkov light from a narrow-band beam of kaons, pions, and protons.  This technique was demonstrated using data taken opportunistically by the Main Injector Particle Production experiment at Fermi National Accelerator Laboratory which recorded beams of protons, kaons, and pions ranging in momentum from +37 GeV/$c$ to +63 GeV/$c$.  The measured value is 491.3 $\pm$ 1.7~MeV/$c^2$, which is within 1.4$\sigma$ of the world average.  An improvement of two orders of magnitude in precision would make this technique useful for resolving the ambiguity in the X-ray data and may be achievable in a dedicated experiment.
\end{abstract}

%\begin{keyword}

%\PACS 
%\end{keyword}
%\tableofcontents
\end{frontmatter}

%%
%% MAIN TEXT
%%

%% Introduction
\section{Introduction}
\label{sec:intro}
The charged kaon mass is an important input in determining the CKM matrix element $V_{us}$ from measurements of the branching ratio of $K^{+} \rightarrow \pi^{0} e^{+} \nu$. The value of the charged kaon mass reported by the Particle Data Group is 493.677~MeV/$c^2$ with an uncertainty of 26 parts per million (ppm)~\cite{eid:pdg}.  This value is a weighted average of six measurements but is dominated by the two most recent and precise measurements from Denisov \cite{den:kmass} and Gall \cite{gall:kmass} which measure X-ray energies from kaonic atoms. While these measurements report uncertainties of 14 and 22~ppm they differ by 122 ppm (4.6$\sigma$). In this article we explore one possibility to resolve this discrepancy using an independent technique for measuring the charged kaon mass based on the Cherenkov effect. The well known pion and proton masses are used as references. The technique is demonstrated using data taken opportunistically using the Ring Imaging Cherenkov (RICH) sub-detector of the Main Injector Particle Production (MIPP) experiment at Fermilab \cite{mipp:collab}.

%% Concept
\section{Measurement Concept}
\label{sec:concept}
Cherenkov light is emitted when a relativistic charged particle of
mass $m$, momentum $p$, and speed $\beta = 1/\sqrt{1+(m/p)^2}$ travels
through a radiator volume of index of refraction $n$ with $\beta >
1/n$. (As is customary in high energy physics we work in units in which
the speed of light, $c$, is 1.) Neglecting dispersive effects for the
moment, the light is emitted in a cone at an angle $\theta$ given by
$\cos \theta = 1/\beta n$~\cite{zrelov:ckov,jelley:ckov} which is
approximately
\begin{equation}
\theta = \sqrt{2(1 - \frac{1}{n\beta})}
\label{eq:thappr}
\end{equation}
for small angles. Now, consider two particles $i$ and $j$ with
identical momenta $p$ but different masses $m_{i}$, $m_{j}$ and speeds
$\beta_i$, $\beta_j$. They will emit Cherenkov light at angles
$\theta_{i}$, $\theta_{j}$ which are related by the expression
\begin{equation}
\beta_i \theta_i^2 - \beta_j \theta_j^2 = 2(\beta_i - \beta_j).
\label{eq:angrel_beta}
\end{equation}
In the relativistic limit $p \gg m$, $\beta \approx 
1 - \frac{m^{2}}{2p^{2}}$.
This, when combined with Eqn.~\ref{eq:angrel_beta}, gives
\begin{equation}
\theta_i^2 -\theta_j^2 = \frac{m_j^2 - m_i^2}{p^2},
\end{equation}
where we have neglected the small difference between $\beta \theta$
and $\theta$.  If the particles $i$ and $j$ are pions, protons, and
kaons, we have two independent angle-squared differences that can be
measured
\begin{equation}
\theta_{\pi}^{2} - \theta_{K}^{2} = \frac{m_{K}^{2} - m_{\pi}^{2}}{p^{2}},
~{\rm and}~~
%\end{equation} 
%\begin{equation}
\theta_{\pi}^{2} - \theta_{p}^{2} = \frac{m_{p}^{2} - m_{\pi}^{2}}{p^{2}}.
\end{equation}
Using these, the kaon mass is given by
\begin{equation}
m_{K}^{2} = m_{\pi}^{2} + 
\Delta_{p\pi}
\frac{\theta_{\pi}^{2} - \theta_{K}^{2}}{\theta_{\pi}^{2} - \theta_{p}^{2}},
\label{eq:kmass}
\end{equation}
where $\Delta_{p\pi} \equiv m_{p}^{2} - m_{\pi}^{2}$. Notice that for
a monochromatic beam in the absence of dispersion the index of
refraction $n$ and momenta $p$ drop out. The kaon mass can be determined, in
principle, through measurements of the pion and proton masses and the
Cherenkov angles of the three particles. The proton and pion
masses are known to 0.9~ppm and 2.5~ppm respectively
and will not be the limiting factors in the experiment.

% \subsection{Estimated Uncertainty}
% \label{subsec:est_unc}

Using Eqn.~\ref{eq:kmass} we can estimate the uncertainty in $m^2_K$
measured using this method as:
\begin{eqnarray}
\sigma^{2}_{m^{2}_{K}} &=&
\sigma^2_{m^2_\pi} + 
\left( 
\frac{\theta^2_\pi - \theta^2_K}{\theta^2_\pi - \theta^2_p} 
\right)^2 \sigma^2_{\Delta_{p\pi}} + \nonumber \\
&&
4p^{4}
\left[ 
\theta^{2}_{\pi}
\frac{\Delta^2_{pK}}{\Delta^2_{p\pi}}\sigma^{2}_{\theta_{\pi}} + 
\theta^{2}_{K}\sigma^{2}_{\theta_{K}} +
\theta^{2}_{p}
\frac{\Delta^2_{K\pi}}{\Delta^2_{p\pi}}\sigma^{2}_{\theta_{p}} 
\right].
\label{eq:mkerr}
\end{eqnarray}
The first two terms are due to the uncertainties in the pion and
proton masses and are small. The third term grows with momentum and
suggests that it is best to conduct the measurement at as low a
momentum as possible where the differences in the Cherenkov angles are
largest, while staying above proton threshold.

In a RICH detector, the angle $\theta$ can be determined on a
track-by-track basis from the pattern of Cherenkov photons
recorded. However, the light for a single ring will be distributed
about the central angle $\theta$ due to the variation of the index of
refraction of the radiator medium over the wavelengths at which
Cherenkov photons are produced. This gives as contribution to the uncertainty
in the average angle $\theta$ determined from a single track of
\begin{equation}
\sigma'^2_{\theta_{i}} 
= 
\frac{1}{N_{h}}
\left(
\frac{1}{\theta_{i}n^{2}\beta_{i}}
\right)^{2}\delta^{2}_{n},
\label{eq:ring_width}
\end{equation}
where $\delta_n$ is the amount of dispersion
over the photomultiplier tube (PMT) wavelength acceptance and
$N_{h}$ is the number of PMT hits in the ring.

\begin{figure}[htbp]
\begin{center}
\includegraphics[width=2.75in]{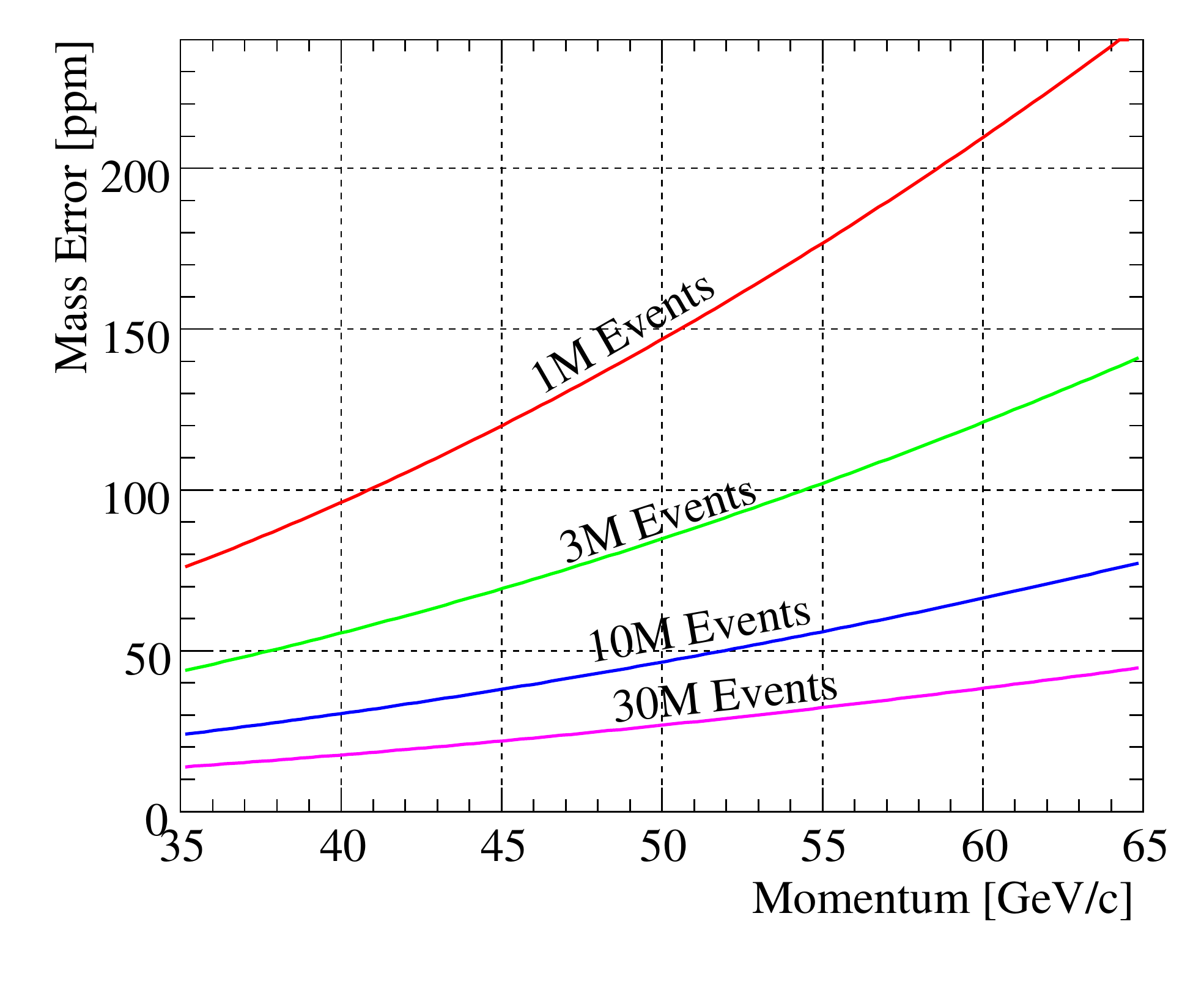}
\caption[Expected statistical uncertainty for kaon mass measurement]
{ Expected statistical uncertainty for kaon mass measurement. }
\label{fig:staterr}
\end{center}
\end{figure}

In a beamline, particles will be accepted if their momentum lies in a
narrow window about some central value $p$. The finite size of this
momentum acceptance window introduces an additional uncertainty in the
average angle measured from a single track. Averaging over $N_r$
rings, the momentum spread contributes an uncertainty
\begin{equation}
\sigma^{2}_{\theta_{i}} = 
\frac{1}{N_r}
\left[
\sigma'^2_{\theta_{i}} +
\left(
\frac{m^{2}_{i}\beta_{i}}{\theta_{i}np^2}
\right)^{2}
(\sigma_{p}/p)^{2}
\right]
\end{equation}
to the measurement of $\theta$, where $\sigma_p$ is the spread in the
beam particle momenta about their central value.  We take the specific
case of an experiment using CO$_2$ as the radiator ($n=1.00045$,
$\delta_n = 3 \times 10^{-5}$) and a beam of central momentum 40~GeV/$c$
with a width of $\sigma_p/p$=0.01. Under these assumptions, 
$\sigma_{\theta_{i}}$ values are in good agreement with observed 
widths of ring radius distributions. 
This gives a statistical uncertainty in the kaon mass of
\begin{equation}
\frac{\sigma_{m_{K}}^2}{m_K^2} =
0.6^2 + 
0.9^2 + 
\frac{31.0^2}{N_{\pi}} +
\frac{44.1^2}{N_{K}} +
\frac{18.1^2}{N_{p}}
[{\rm ppm}]^2,
\label{eq:statfinal}
\end{equation}
where the first term results from uncertainties in the pion mass, the
second from uncertainties in the proton mass, and the final three
terms result from uncertainties in the angles $\theta_\pi$,
$\theta_K$, and $\theta_p$ with $N_{\pi}$, $N_{K}$, and $N_{p}$ being
the number of millions of pion, kaon, and proton rings recorded. Using
Eqn.~\ref{eq:statfinal} we find that the uncertainty is minimized if
32\% of the data is collected using protons, 23\% using pions, and
45\% using kaons. This result is only weakly dependent on momentum as
shown in Figure~\ref{fig:staterr} which plots the expected statistical
precision of the charged kaon mass as a function of momentum choice
and total number of rings recorded.
With 10 million rings at 40 GeV/c, we expect a statistical precision of
30~ppm using this technique.

%% RICH
\section{RICH detector overview}
\label{sec:rich}
%\subsection{Physical Description}
%\label{subsec:rich_desc}

The RICH detector used by MIPP was built by the SELEX Collaboration \cite{selex:collab} for use in that experiment. We summarize here only the most important features of the detector as deployed for the MIPP experiment and refer the reader to \cite{selex:rich1,selex:rich2,selex:rich3,selex:ronchi} for details.

\begin{figure}[htbp]
\begin{center}
\includegraphics[width=3.4in]{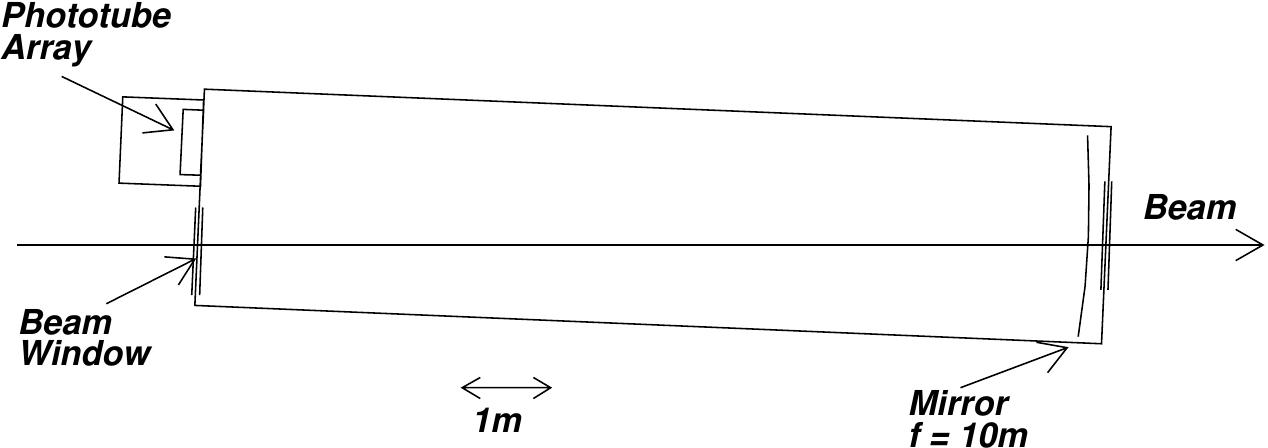}
\caption
{ Schematic of the RICH detector \cite{selex:rich1}. }
\label{fig:rich_sch}
\end{center}
\end{figure}
The geometry of the RICH counter is shown in Figure~\ref{fig:rich_sch}. The detector was constructed from a low carbon cylindrical steel vessel 10.22~m in length and 93~in. in diameter with a wall thickness of $\frac{1}{2}$~in. The ends were sealed with 1.5~in. thick aluminum flanges that were cut out to hold thin beam windows at each end and a photomultiplier tube holder plate at the upstream end. The vessel was wrapped by a water line carrying chilled water and 15~cm of building insulation to regulate the temperature.

A 2.4~m$\times$1.2~m array of 16 mirrors mounted at the downstream end of the counter focused Cherenkov light on an array of 1/2~in. photomultiplier tubes. On average the mirrors had a radius of curvature of 1980~cm with variations less than 5~cm and a reflectivity of more than 85\% at 160~nm.

\begin{figure}[htbp]
\begin{center}
\includegraphics[width=2.5in]{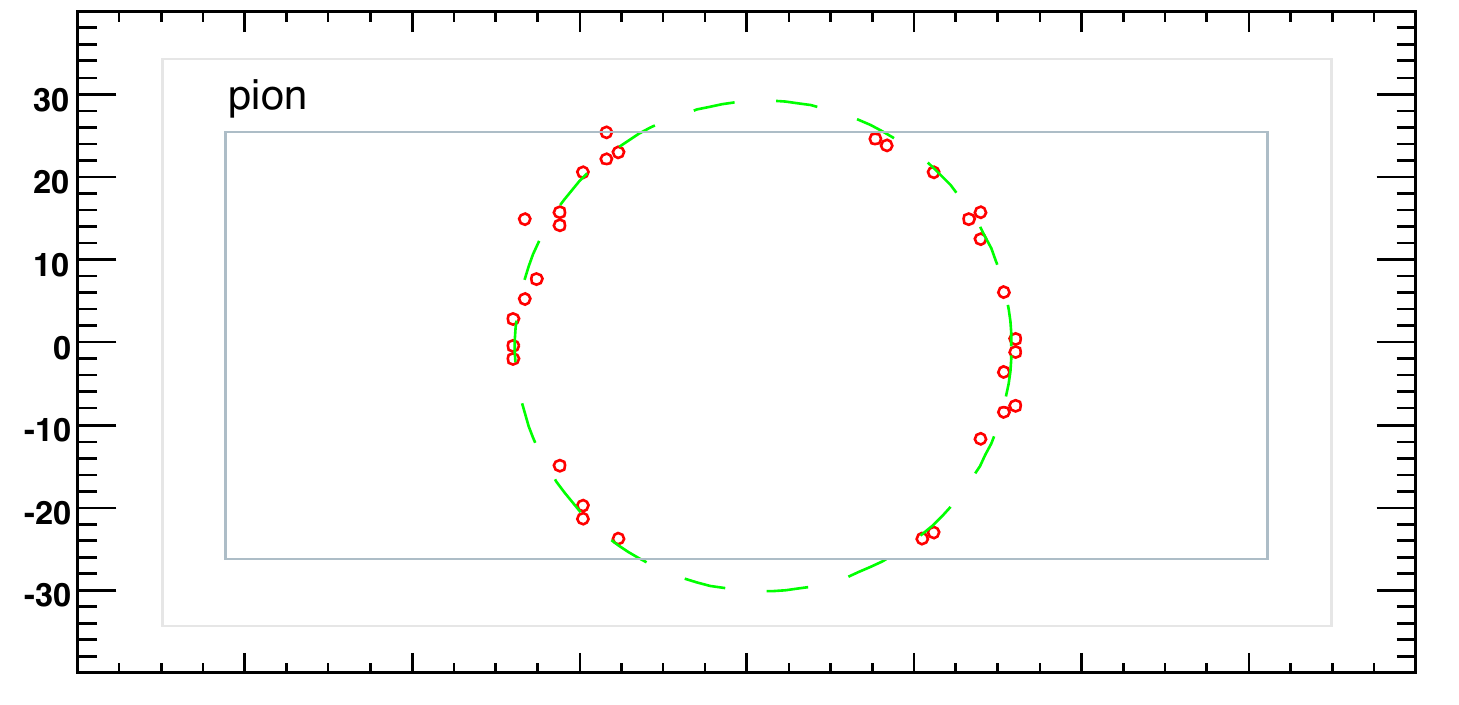}
\includegraphics[width=2.5in]{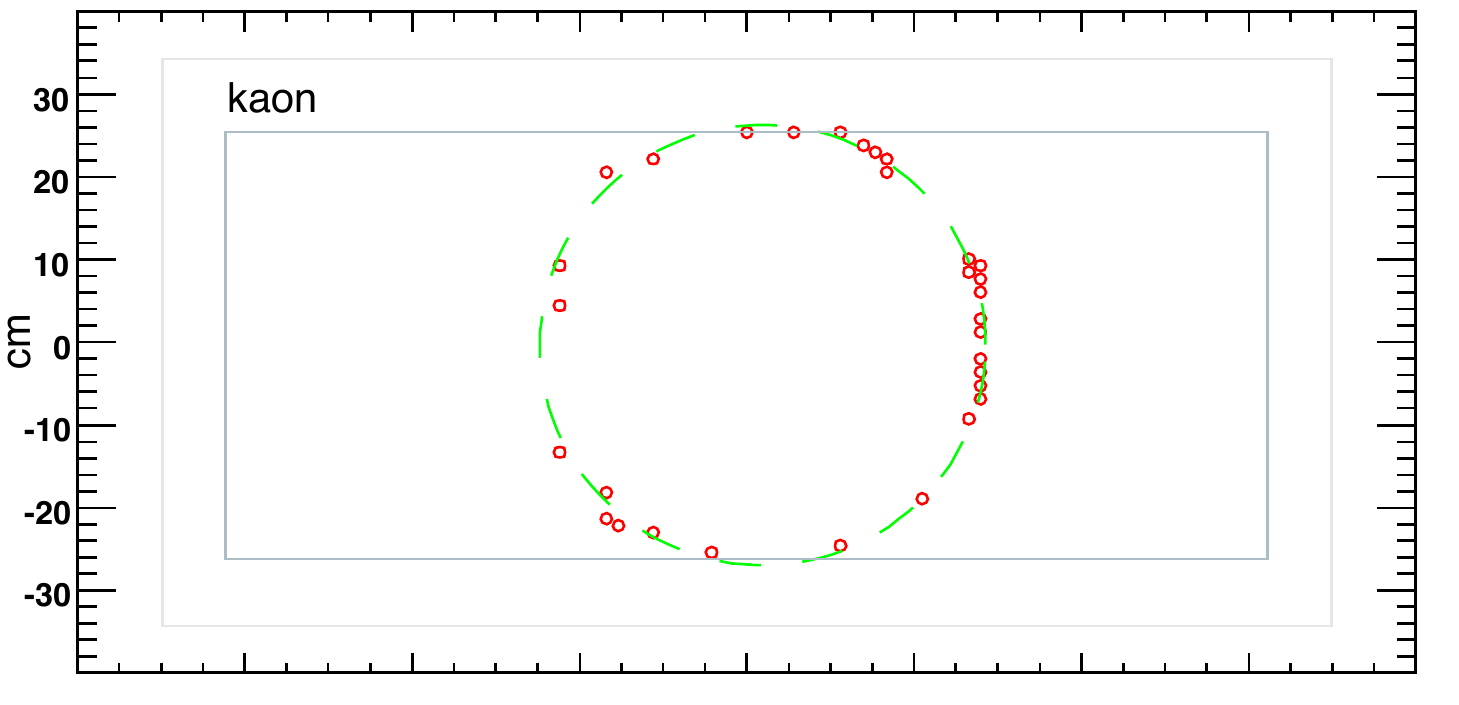}
\includegraphics[width=2.5in]{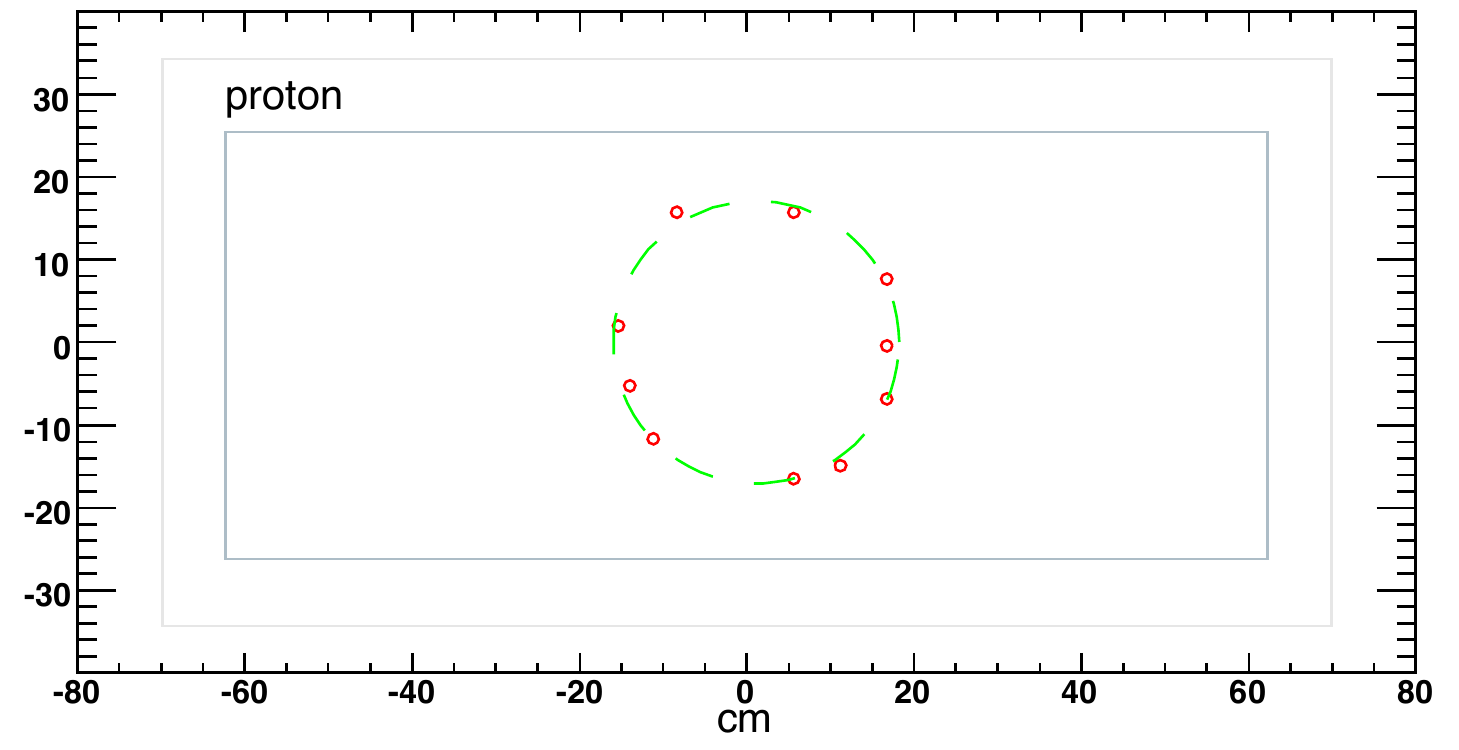}
\includegraphics[width=2.6in]{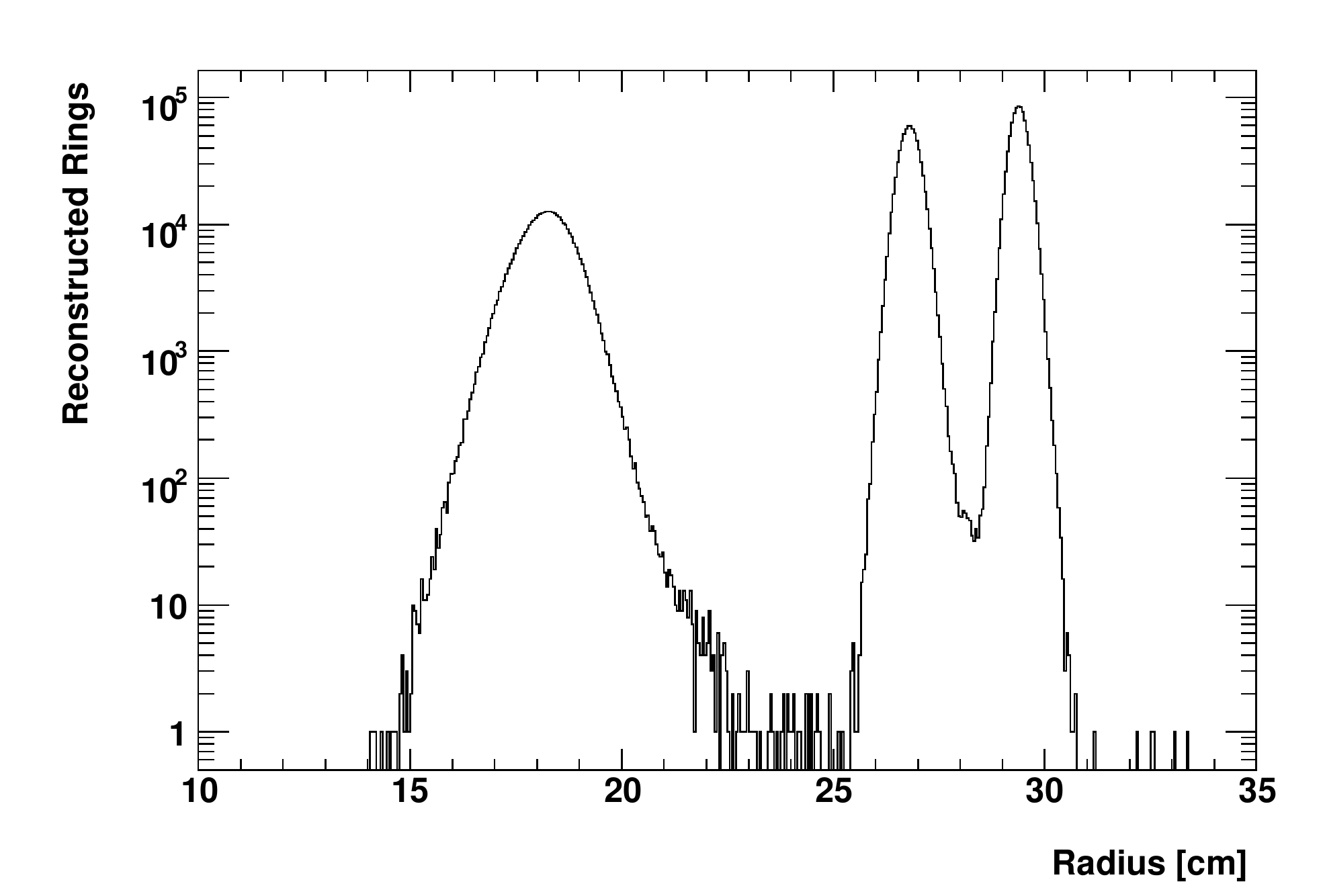}
\caption{ Sample event displays of 40~GeV pion (top), kaon (second from top), and proton (second from bottom) rings in the RICH counter. Small circles indicate hit PMTs. The large dashed circles show the rings reconstructed from the PMT hits.  The bottom plot is the distribution of reconstructed ring radii for 40~GeV.}
\label{fig:evd}
\end{center}
\end{figure}

MIPP used CO$_{2}$ gas held at just above atmospheric pressure as the radiator. At STP, CO$_2$ has an index of refraction of 1.00045 at $\lambda=300$~nm giving thresholds of $p=4.5$~GeV/$c$ for pions, 17~GeV/$c$ for kaons, and 31~GeV/$c$ for protons. The gas temperature and pressure were continuously monitored enabling calibration of the index of refraction. A $\beta = 1$ particle produced a ring of radius 29.5~cm and an average of 30 PMT hits providing 3$\sigma$ $\pi$/$K$ separation up to 80 GeV/$c$ and 3$\sigma$ $p$/$K$ separation up to 120 GeV/$c$.  Figure~\ref{fig:evd} shows sample event displays for each beam species showing PMT hits and reconstructed rings. 

% \subsection{Phototubes}
% \label{subsec:rich_pmt}

The PMTs were mounted in a hexagonal array behind a 2~mm thick quartz window which provided a gas-tight seal between the phototubes and the radiator volume. They were read out in threshold (on/off) mode.  Two models of PMTs were used: Hamamatsu R760 and Russian made FEU60. The FEU60 tubes were coated with a wavelength shifter to match the acceptance region of the R760's. The R760 (FEU60) tubes have a maximum quantum efficiency of about 
25\% (11\%) at 350~nm. Of the available 89 PMT columns, 68 were used, with R760's installed in 15 columns and FEU60's installed in 53 columns.  The front-end electronics of the RICH detector was re-designed and rebuilt by the MIPP collaboration.

%% Data Analysis
\section{Data Analysis}
\label{sec:ana}
In outlining the measurement concept, we made several simplifying
assumptions which must be accounted for during analysis. Due to
dispersion, Cherenkov light is not observed at a single angle, but as
a distribution across several angles. In our analysis, we measured the
average PMT occupancies for pion, proton, and kaon rings as a function
of Cherenkov angle in data and compared them to calculations which incorporated
the kaon mass as a free parameter.  In total 12 million rings were recorded
for this measurement using positively charged beams of pions, kaons, and
protons ranging from 37 GeV/$c$ to 63 GeV/$c$ in momentum.

In the data, wire chambers upstream and downstream of the RICH counter
were used to reconstruct the particle trajectory and predict the
center of the RICH ring. Using that prediction for the center
position, a Cherenkov angle was assigned to each PMT and the average
probability for a PMT to fire was computed as a function of Cherenkov
angle. This procedure was done
separately for each PMT type, mirror, and momentum setting. As the
measurement was made with beam particles, light struck only the two
central mirrors, labeled 8 and 9.

\begin{figure}[htbp]
\begin{center}
\includegraphics[width=2.5in]{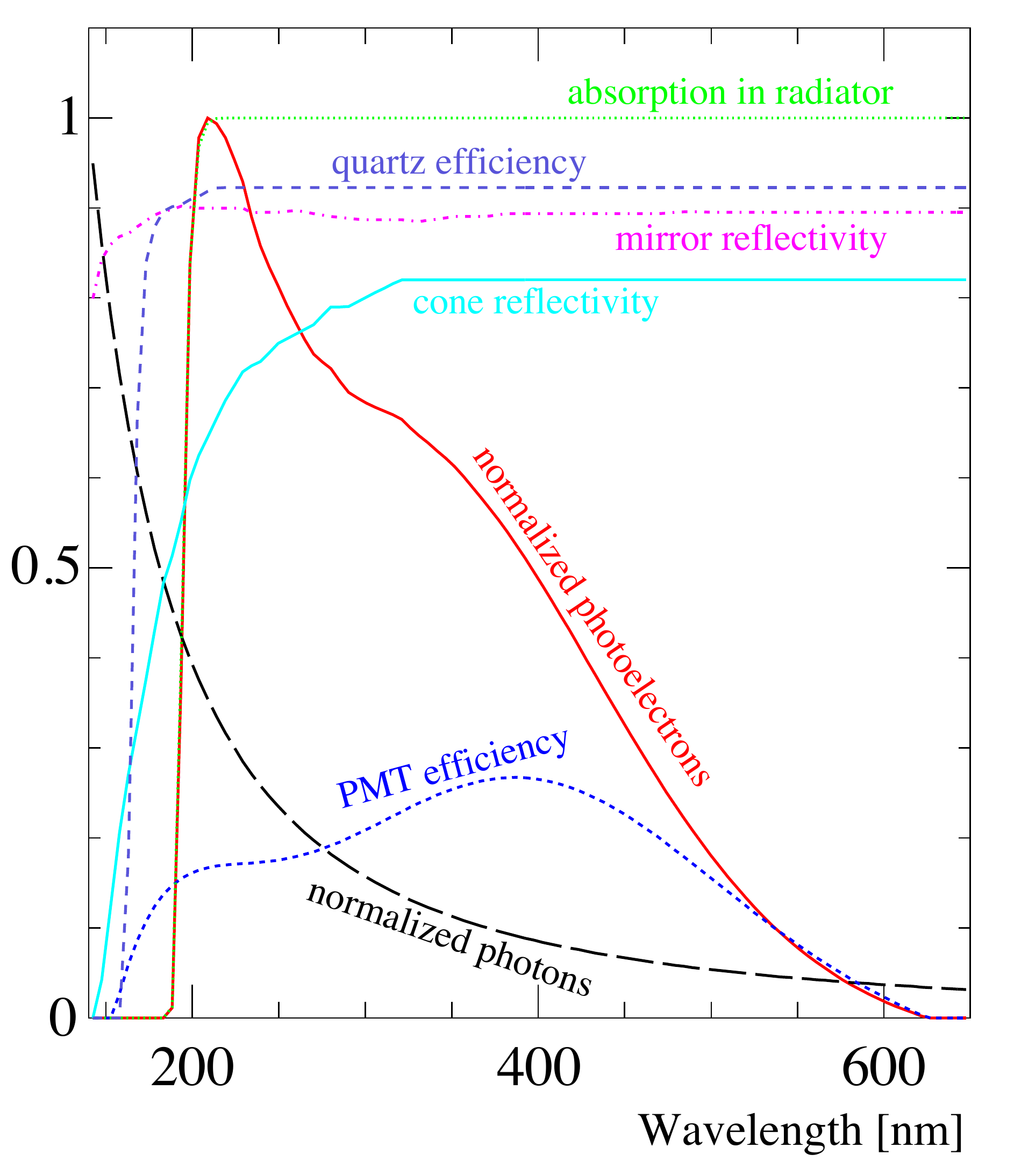}
\caption
{ Overlay of all efficiency functions incorporated into the
  calculation of detector photoelectrons as a function of angle. Also
  shown are the initial Cherenkov photon production spectrum and final
  photoelectron yield normalized to peak at 1 for inclusion in the
  plot.  }
\label{fig:light_shape}
\end{center}
\end{figure}

To compare to the data, we calculated the expected occupancy of each
PMT in the RICH array for pion, proton, and kaon rings. The
calculation starts with the number of Cherenkov photons produced per
unit path length between wavelengths $\lambda$ and $\lambda +
d\lambda$~\cite{zrelov:ckov,jelley:ckov}:
\begin{equation}
\frac{d^{2} N_{ph}}{d \lambda dx} = \frac{2 \pi \alpha}{\lambda^{2}} \sin^{2} \theta_C,
\label{eq:lightyield}
\end{equation}
where $\alpha$ is the fine structure constant. These photons travel
through the radiator medium, reflect off mirrors, pass through a
quartz window, and are collected by a reflective cone before they are
incident on the photodetector. The transmission probability for a
Cherenkov photon of wavelength $\lambda$ for each of these steps is
plotted in Figure~\ref{fig:light_shape} along with the photodetector
efficiency. Combining these factors, the average number of
photoelectrons detected by PMT $i$ is given by
\begin{eqnarray}
N^{i}_{pe} &=& 
\int^{L}_{0}
\int^{\theta_2}_{\theta_1} 
\int_{\lambda_{1}}^{\lambda_{2}}
\frac{2 \pi \alpha}{\lambda^{2}} 
\left(
1 - \frac{1}{n^{2}(\lambda)\beta^{2}}
\right)
e^{-\mu(\lambda)(F_{L}+x)} \nonumber \otimes \\
&&
S(\theta,\theta_C(\lambda)) \epsilon(\lambda) G_{i}(\theta)
d\lambda d\theta dx
\label{eq:pmt_npesmear}
\end{eqnarray}
where $\epsilon(\lambda)$ is the product of all wavelength-dependent
efficiency factors, $\mu(\lambda)$ is the absorption coefficient of
CO$_{2}$, $F_{L}$ is the mirror focal length, and $G_i$ is the
geometric acceptance of the $i^{\rm th}$ PMT. Scattering of light from
angular bin $\theta_C$ to bin $\theta$ is accounted for by the
function $S(\theta,\theta_C)$. The light scattering was modeled as a
Gaussian of width $\sigma$ with three components.  These include an
intrinsic term independent of wavelength and a wavelength-dependent
dispersive term~\cite{forty:lhcb}. A third term accounts for
multiple scattering of the beam particle in the RICH radiator:
\begin{equation}
\sigma^2 = \sigma_0^2 + \frac{\sigma_N^2}{\tan^2 \theta_C} + \sigma^2_{ms}
\end{equation}
where $\sigma_0$ is the intrinsic scattering width and $\sigma_n$ is 
the dispersive scattering width.  The calculated shapes are implicitly 
functions of the particle masses through $\beta$.
Using Poisson statistics, the probability for a PMT seeing $N_{pe}$
photoelectrons to be on is:
\begin{equation}
P = 1 - \exp(-N_{pe} - b),
\label{eq:pmt_prob}
\end{equation}
where $b$ accounts for the PMT dark noise rate.

The measured and calculated PMT occupancies were compared using a
$\chi^{2}$ statistic. In the comparison several parameters were
allowed to vary to minimize $\chi^2$. These were:
\begin{enumerate}[i.]
\item Intrinsic detector smearing width, $\sigma_{0}$
\item Dispersive smearing width, $\sigma_{N}$
\item Density ratio scaling factor
\item Density ratio offset
\item Level of air contamination in CO$_{2}$
\item Central pion momentum
\item Width of momentum distribution
\item Kaon mass
\end{enumerate}
The density ratio is defined as the gas density inside the RICH
divided by density at STP and was varied to account for uncertainties
in the calibration of the pressure and temperature monitors installed
in the counter.

The momentum acceptance of the beamline was modeled as a Gaussian and
uncertainties in the mean and width of this acceptance were
incorporated into the calculation. The average proton and kaon momenta
accepted by the beamline differ slightly from the average pion
momentum due to variations in particle production at the secondary
target. This effect was incorporated using a simulation of the copper
secondary target using the FLUKA~\cite{fluka1,fluka2} production
model.

\begin{figure}[htbp]
\begin{center}
\includegraphics[width=3.4in]{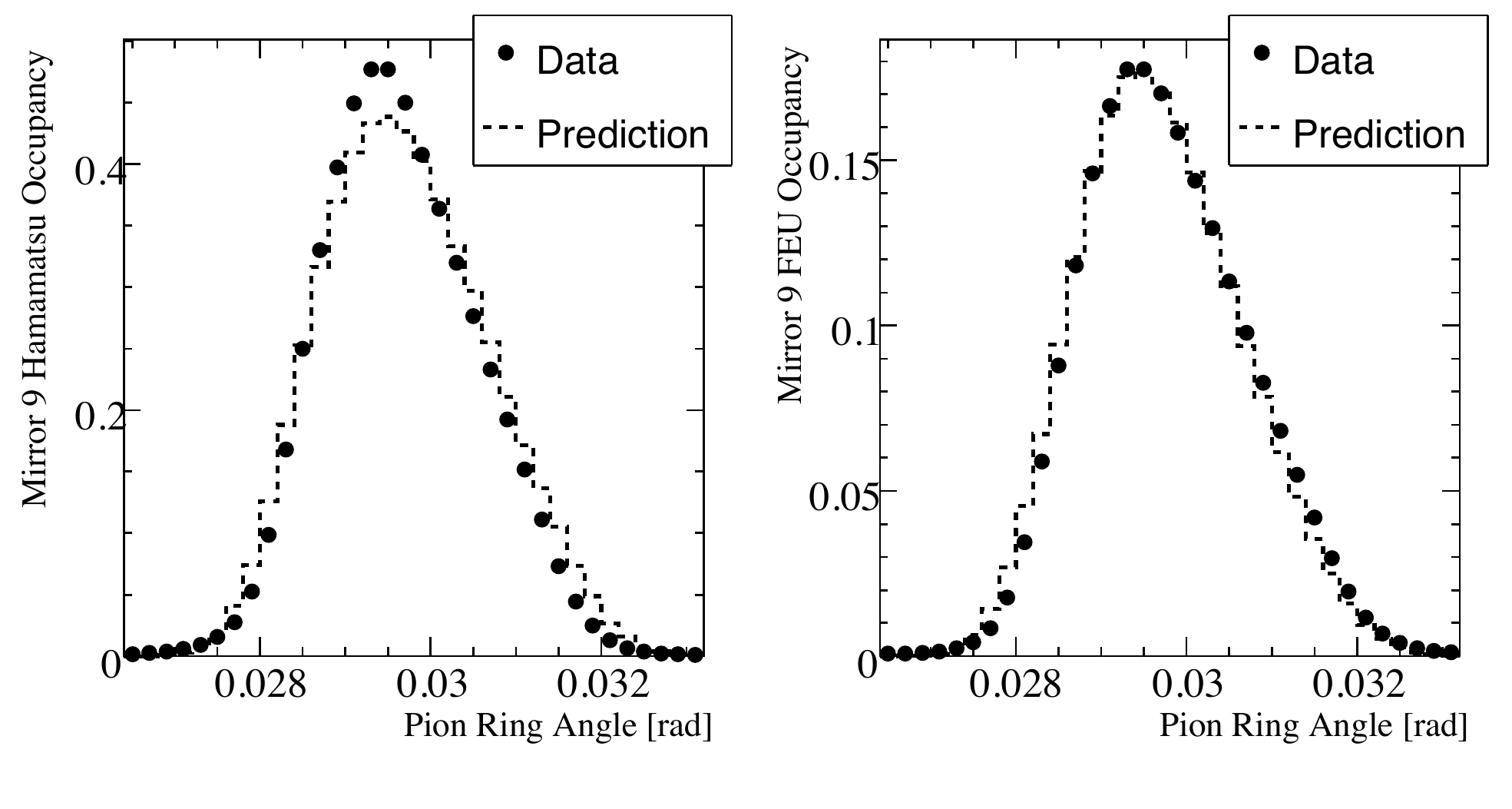}
\includegraphics[width=3.4in]{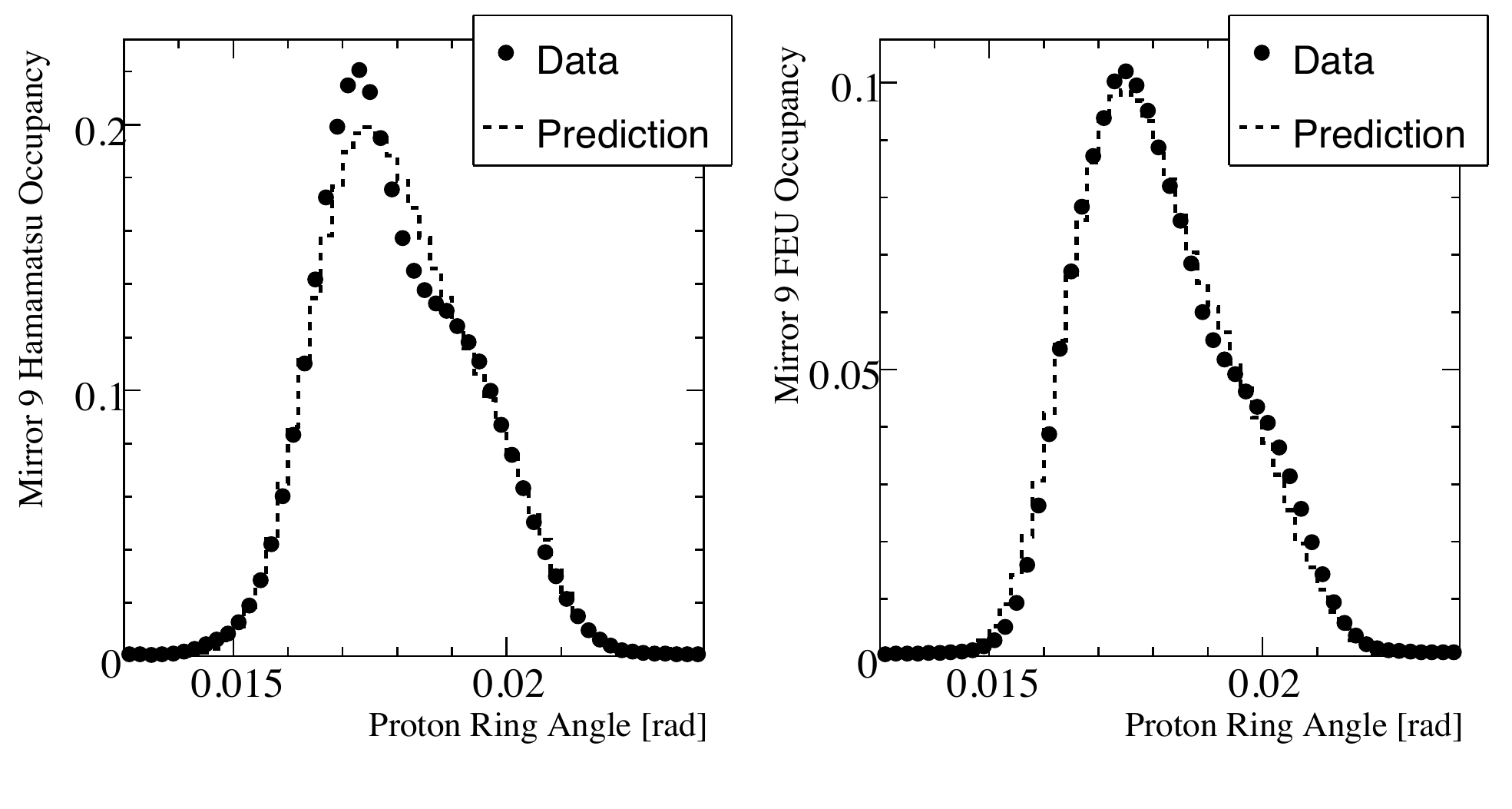}
\includegraphics[width=3.4in]{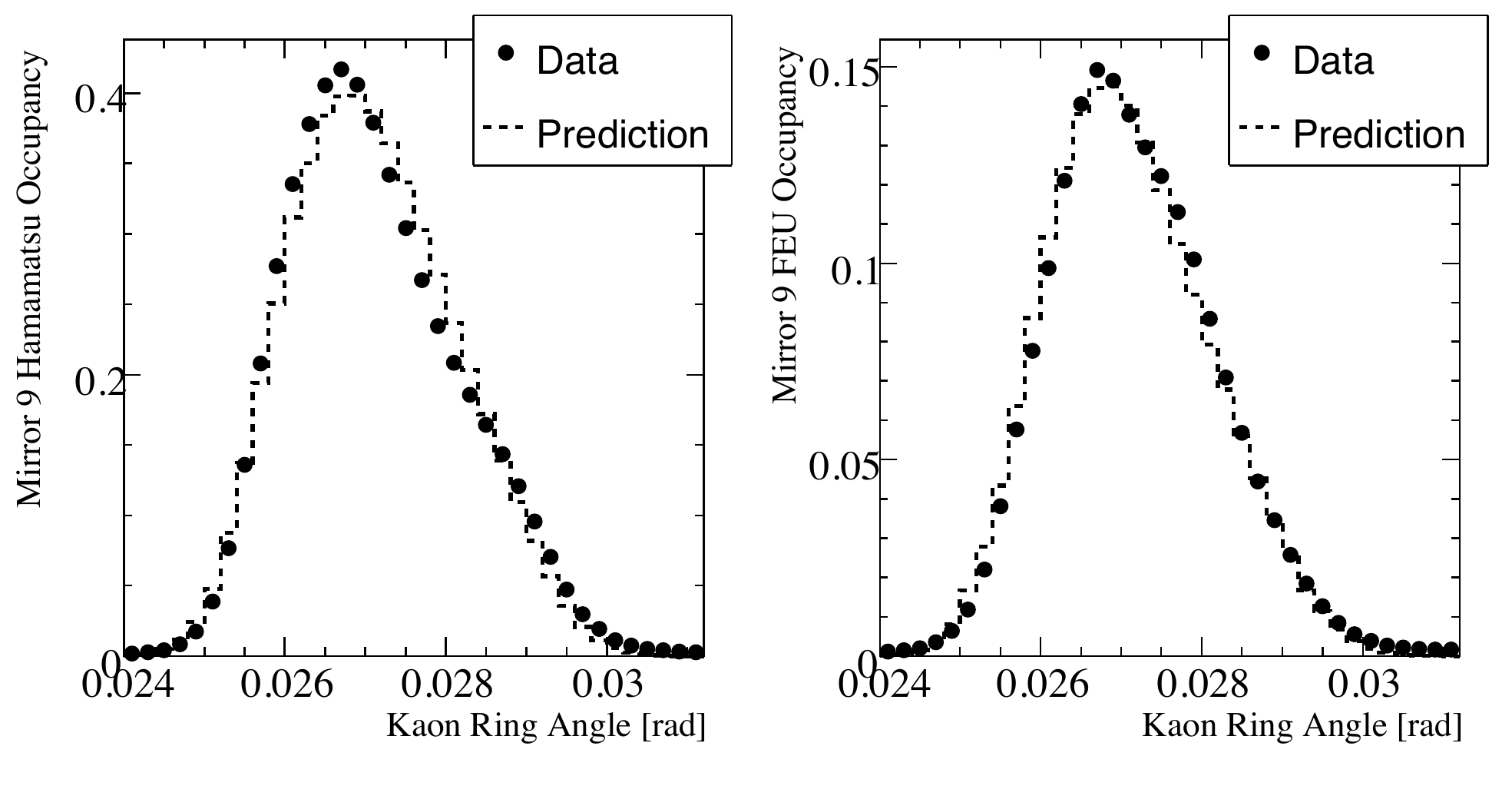}
\caption { Comparison of measured PMT occupancies to best-fit
  predictions for 40 GeV data set~\cite{graf:mipp}.  Pions are shown
  in the top row, protons in the middle row, and kaons in the bottom
  row. In each row, the left panel shows the results for R760 PMTs
  while the right panel shows the results for FEU60 PMTs.  }
\label{fig:occ_40gev}
\end{center}
\end{figure}

\begin{figure}[htbp]
\begin{center}
\includegraphics[width=3.4in]{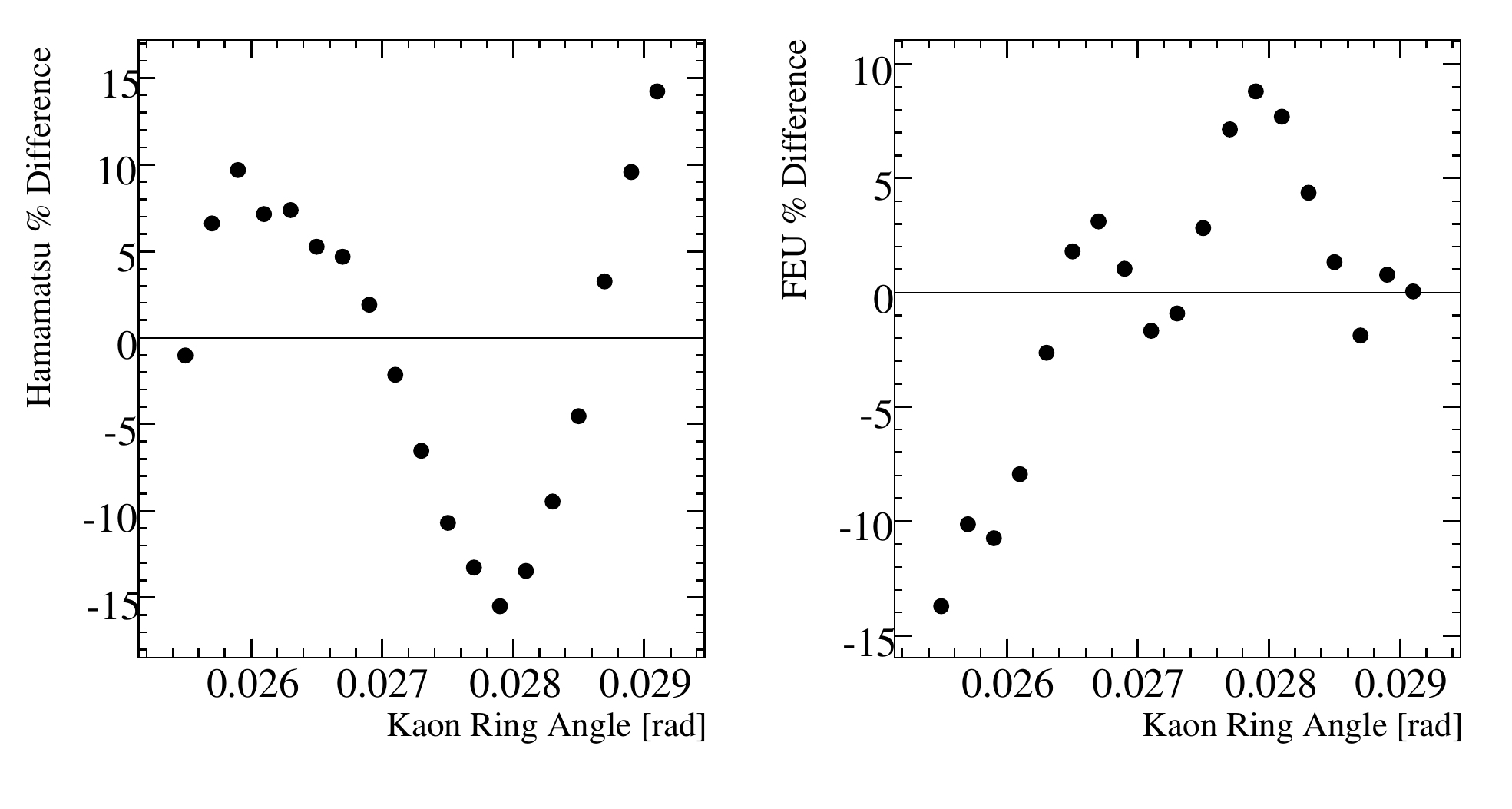}
\caption { Percent difference between data and predicted occupancy
  for 40 GeV/$c$ kaons.  Hamamatsu tubes are on the left, FEU tubes
  are on the right.  }
\label{fig:kaon_resid}
\end{center}
\end{figure}

After all known effects were accounted for there were
$\mathcal{O}$(10\%) differences between the measured and
calculated PMT occupancies, due presumably to our incomplete
understanding of the factors which control Cherenkov photon production
and transport in our detector. To account for these differences, the
uncertainties in the PMT occupancies were increased until $\chi^{2} /
ndf = 1$ was achieved at the minimum. The scaling factor ranged from
about 25 to 100 for the different combinations of PMT type, mirror,
and beam momentum. Systematic uncertainties overwhelm the
statistical uncertainties for this data set. A sample fit for 40
GeV/$c$ data is shown in Figure \ref{fig:occ_40gev}.  Percent difference
between data and predicted occupancy is shown in Figure \ref{fig:kaon_resid}.
They agree within 5-15\% over the angular rangue illuminated by kaon rings.
This difference is typically 15 times larger than the statistical uncertainty in
each angular bin.

\begin{figure}[htbp]
\begin{center}
\includegraphics[width=3.4in]{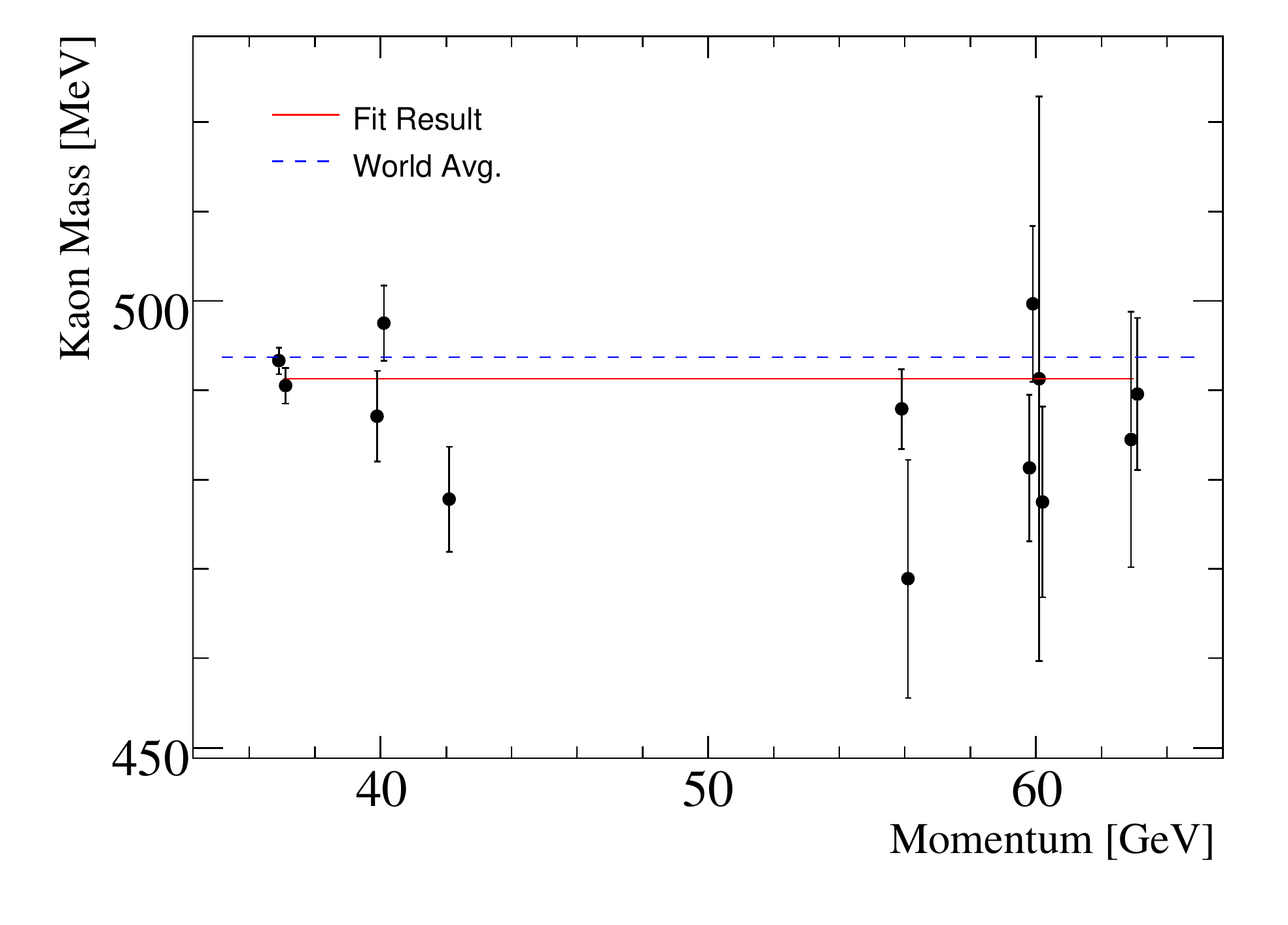}
\caption { Summary of kaon mass results for data sets using various
  beam momenta and mirrors. At each momentum setting, results for
  mirror 8 are plotted shifted slightly to the left, while results for
  mirror 9 are plotted shifted slightly to the right. The solid line
  shows the final kaon mass result, and the dotted line is the
  accepted value from the PDG.  }
\label{fig:final_fit}
\end{center}
\end{figure}

Kaon mass results for each data set are shown in
Figure~\ref{fig:final_fit}. Using just the low momentum data sets
results in a kaon mass measurement of 491.9 $\pm$ 1.1 MeV; using just
the high momentum data sets gives 486.7 $\pm$ 3.0 MeV. These results
agree within uncertainties and are combined to give a final result for
the charged kaon mass of 491.3 $\pm$ 1.7 MeV (3500 ppm).  The
convention of scaling error bars when combining measurements so that
$\chi^{2} / ndf = 1$ is followed resulting in a larger final
uncertainty.  This result is within 1.4$\sigma$ of the PDG
value~\cite{eid:pdg}.

See \cite{graf:mipp} for a detailed discussion of the analysis presented in 
this paper.

%% Improvement
\section{Suggestions for Improvement}
\label{sec:improv}
In the analysis above we achieved an understanding of the PMT
occupancies as a function of Cherenkov angle at the 10\% level
resulting in a final uncertainty in the kaon mass of 3500 ppm.  If the
uncertainties in the PMT occupancies could be reduced to 1\% the
contribution of this source of uncertainty in the kaon mass would be
reduced to the level of the statistical uncertainties and make this
technique useful for resolving the 122~ppm discrepancy in the X-ray
measurements.

The largest uncertainty in our calculation of the PMT occupancy as a
function of angle results from knowledge of the acceptance of the PMT
array which was complicated by cross-talk in the readout electronics
-- a problem which manifested itself during the high rate conditions
the detector was operated under for this measurement.  Occasionally, a
single PMT hit would cause all 16 channels sharing the same readout
board to register a hit.  In our analysis we rejected PMT hits that
appeared to be due to cross-talk.  As the cross-talk was difficult to
model, its affect on the PMT acceptances has large uncertainties
accounting for most of the $\sim$10\% uncertainty in PMT occupancy as
a function of angle.

Additional uncertainties in the occupancies result from our treatment
of the detector response as a function of wavelength.  A measurement
of the PMT response on a tube-by tube basis would have required
dis-assembly of the detector and was not undertaken, rather we used a
single average response for all PMTs.  As the response of individual
PMTs typically differ from the average by about 15\%, this introduces
an uncertainty of $\sim$6\% in the calculation of occupancy versus angle.
Likewise, we assumed identical reflection efficiencies for the two
reflecting mirrors.  Taken together, the uncertainties in the PMT
acceptances and spectral response of the optical system contribute
3400 ppm to our measurement of the kaon mass.

During the run the temperature and pressure of the radiator gas varied
considerably and the index of refraction had to be calibrated several
times a day.  We estimate that uncertainties in the knowledge of index
of refraction contributed 500~ppm in our kaon mass measurement.  The
spread on the beam momentum about its central value contributed
roughly 50~ppm to the kaon mass measurement.

To achieve 40 ppm precision with this technique each of the
uncertainties listed above would need to be addressed.  The largest
improvements are to be gained by using a simpler optical system which
could be completely characterized and operated at a more consistent
pressure and temperature free from electronic cross-talk problems.
For example, the optical system would be simpler if only a single
primary mirror could be used and the transmission windows eliminated.
Use of a fine-grained optical detector would allow for the device to
be made much more compact allowing for better control of pressure and
temperature.  A more compact device could be built to run with large
excursions in gas pressure and temperature allowing for the variation
in the index of refraction with temperature and pressure to be studied
in situ.  Finally, a charge based readout (rather than the on/off
readout used in the above analysis) would allow for better
pixel-by-pixel knowledge of the detector response.  This configuration
could be achieved, for example, by replacing the secondary mirror in
the MIPP beam Cherenkov counter with a GEM-based photo-cathode
\cite{Azevedo:2009em} which would afford 30~micron resolution of the
Cherenkov rings in a device with is 45.7~cm in diameter and 22.9~m
long.  A complete assessment of the performance of such a device would
require a significant research effort and awaits an experimental
proposal.  We hope that this initial work will inspire these future
investigations.

In conclusion, we have used the Cherenkov effect to measure the
charged kaon mass with 3500~ppm precision using an existing RICH
detector.  We believe that there are significant opportunities for a
future experiment dedicated to the kaon mass measurement to reduce
systematic uncertainties to the level where this technique may be
useful to resolve the 122~ppm discrepancy in the X-ray measurements of
the charged kaon mass.

%% Acknowledgments
\section*{Acknowledgments}
Fermilab is operated by Fermi Research Alliance, LLC under Contract No. DE-AC02-07CH11359 with the United States Department of Energy. 

\bibliographystyle{utphys}
\bibliography{kmass_nim}

\providecommand{\href}[2]{#2}\begingroup\raggedright\begin{thebibliography}{10}

\bibitem{eid:pdg}
C.~Amsler {\em et al.}, ``Particle data group,'' {\em Phys. Lett. B667} {\bf 1}
  (2008)  704.

\bibitem{den:kmass}
A.~Denisov {\em et al.}, ``New measurement of the mass of the ${K}^{-}$
  meson,'' {\em JEPT Lett.} {\bf 54} (1991)  558.

\bibitem{gall:kmass}
K.~Gall {\em et al.}, ``Precision measurement of the ${K}^{-}$ and $\sigma^{-}$
  masses,'' {\em Phys. Rev. Lett.} {\bf 60} (1988)  186.

\bibitem{mipp:collab}
``{MIPP} {C}ollaboration.''. \url{http://ppd.fnal.gov/experiments/e907/}.

\bibitem{zrelov:ckov}
V.~Zrelov, {\em Cherenkov Radiation in High-Energy Physics}.
\newblock Israel Program for Scientific Translations, Ltd., 1970.

\bibitem{jelley:ckov}
J.~Jelley, {\em Cherenkov Radiation and its Applications}.
\newblock Pergamon Press, 1958.

\bibitem{selex:collab}
``{SELEX} {C}ollaboration.''. \url{http://fn781a.fnal.gov}.

\bibitem{selex:rich1}
J.~Engelfried {\em et al.}, ``The {SELEX} phototube {RICH} detector,'' {\em
  Nucl. Instr. and Meth. A} {\bf 431} (1999)  53--69.

\bibitem{selex:rich2}
J.~Engelfried {\em et al.}, ``The {RICH} detector of the {SELEX} experiment,''
  {\em Nucl. Instr. and Meth. A} {\bf 433} (1999)  149--152.

\bibitem{selex:rich3}
J.~Engelfried {\em et al.}, ``The e781 ({SELEX}) {RICH} detector,'' {\em Nucl.
  Instr. and Meth. A} {\bf 409} (1998)  439--442.

\bibitem{selex:ronchi}
L.~Stutte, J.~Engelfried, and J.~Kilmer, ``A method to evaluate mirrors for
  {C}herenkov counters,'' {\em Nucl. Instr. and Meth. A} {\bf 369} (1996)  69.

\bibitem{forty:lhcb}
R.~Forty and O.~Scheider, ``{\it {RICH} pattern recognition}.'' Lhcb technical
  note number lhcb/98 040, 1998.
\newblock \url{http://lhcb.web.cern.ch/lhcb-rich/html/lhcb_rich_notes.htm}.

\bibitem{fluka1}
A.~Fasso', A.~Ferrari, J.~Ranft, and P.~Sala, ``{FLUKA}: a multi-particle
  transport code.'' Cern 2005-10, infn/tc\_05/11, slac-r-773, 2005.

\bibitem{fluka2}
A.~Fasso' {\em et al.}, ``The physics models of {FLUKA}: status and recent
  developments,'' in {\em Computing in High Energy and Nuclear Physics
  Conference}.
\newblock 2003.

\bibitem{graf:mipp}
N.~J. Graf, {\em Measurement of the Charged Kaon Mass with the {MIPP} {RICH}}.
\newblock PhD thesis, Indiana University, Bloomington, IN, August, 2008.

\bibitem{Azevedo:2009em}
C.~D.~R. Azevedo {\em et al.}, ``{Towards THGEM UV-photon detectors for RICH:
  on sinlge-photon detection efficiency in Ne/CH4 and Ne/CF4},''
\href{http://arxiv.org/abs/0909.5357}{{\tt arXiv:0909.5357 [physics.ins-det]}}.
%%CITATION = 0909.5357;%%.

\end{thebibliography}\endgroup

\end{document}